\documentclass[aps,prb,groupedaddress,showpacs,twocolumn]{revtex4}
\usepackage{amssymb}
\usepackage{bm}
\usepackage{graphicx}
\usepackage{color}

\begin{document}

\title{Domain wall superconductivity in superconductor/ferromagnet bilayers}
\author{M. Houzet$^1$ and A. I. Buzdin$^2$}
\affiliation{
$^1$ DRFMC/SPSMS, CEA Grenoble, 17, rue des Martyrs, 38054 Grenoble cedex 9,
France\\
$^2$ Institut Universitaire de France and Universit\'{e} Bordeaux I, CPMOH, UMR
5798, 33405 Talence, France }
\date{\today}

\begin{abstract}
We analyze the enhancement of the superconducting critical temperature of 
superconducting/ferromagnetic bilayers due to the appearance of localized 
superconducting states in the vicinity of magnetic domain walls in the ferromagnet. 
We consider the case when the main mechanism of the superconductivity destruction via 
the proximity effect is the exchange field. We demonstrate that the
influence of the domain walls on the superconducting properties of the bilayer
may be quite strong if the domain wall thickness is of the order of
superconducting coherence length.
\end{abstract}


\maketitle

\section{Introduction}

The coexistence of singlet superconductivity and ferromagnetism is very
improbable in bulk compounds but may be easily achieved in artificially
fabricated hybrid superconductor (S)-ferromagnet (F) structures.

There are two basic mechanisms responsible for interaction of
superconducting order parameter with magnetic moments in the ferromagnet:
the electromagnetic mechanism (interaction of Cooper pairs with magnetic
field induced by magnetic moments) and the exchange interaction of magnetic
moments with electrons in Cooper pairs. The second mechanism enters into
play due to the proximity effect, when the Cooper pairs penetrate into the F
layer and induce superconductivity there. In S/F bilayers it is possible to
study the interplay between superconductivity and magnetism in a controlled
manner, since we can change the relative strength of two competing orderings
by varying the layer thicknesses and magnetic content of F layers.
Naturally, to observe the influence of the ferromagnetism on the
superconductivity, the thickness of S layer must be small. This influence is
most pronounced if the S layer thickness is smaller than the superconducting
coherence length $\xi _s$. Recently, the observation of many interesting
effects in S/F systems became possible due to the great progress in the
preparation of high-quality hybrid F/S systems - see the reviews\cite%
{Buzdin-rew,BergEfVolk-rew,Lyuksutov}.

In practice, the domains appear in ferromagnets and, near the domain walls, a
special situation occurs for proximity effect. For the purely orbital
(electromagnetic) mechanism of the superconductivity destruction, the
nucleation of the superconductivity in the presence of domain structure has
been theoretically studied in \cite{BuzMel,Aladyshkinetal} for the case of
magnetic film with perpendicular anisotropy. The conditions of the
superconductivity appearance occur to be more favorable near the domain
walls due to the partial compensation of the magnetic induction. Recently,
the manifestation of such domain wall superconductivity (DWS) was revealed
on experiment \cite{Yangetal} where Nb film was deposited on top of 
the single crystal ferromagnetic BaFe$_{12}$O$_{19}$ covered with a thin Si buffer layer.

As the typical value of the exchange field in the ferromagnets $h\sim
(100-1000)K$ exceeds many times the superconducting critical temperature $%
T_{c0}$, the exchange mechanism prevails the orbital one in the
superconductivity destruction when the electrical contact between S and F
layers is good. For the proximity effect mediated by the exchange
interaction, the Cooper pairs feel the exchange field averaged over the
superconducting coherence length. Naturally, it will be smaller near the
domain wall and we may expect that superconductivity would be more robust
near them. The local increase of the critical temperature in presence of 
magnetic domains was observed experimentally in Ni$_{0.80}$Fe$_{0.20}$/Nb
bilayers (with Nb thickness around 20 nm) \cite{Rusanovetal}, and it was 
attributed to DWS formation.

In the present paper we study theoretically the conditions of the localized
superconductivity appearance near the domain wall taking into account
exchange mechanism of the proximity effect. In Sec. \ref{sec2}, we
demonstrate that, in the case of thin F layer and small domain wall
thickness, the problem is somewhat similar to that of the domain wall
superconductivity in ferromagnetic superconductor \cite{BuzBulPan,magsuper}.
For the case when the superconducting coherence length $\xi_s$ 
exceeds the DW thickness $w$, we expect a very strong local increase of $T_{c}$ (see Sec. %
\ref{sec3}). In Sec. \ref{sec4}, we obtain the analytical expression for the
critical temperature of the DWS for the case when the DW thickness$\
w$ exceeds the superconducting coherence length $\xi_s$. The Appendix
presents an extension of this result to arbitrary thickness of the F layers
and transparency of the S/F interface. We discuss our results in Sec. \ref%
{sec5}. In particular, we predict the realization of the situation when
superconductivity appears only near the DW.

\section{Equation for the critical temperature in thin bilayers}

\label{sec2}

We introduce the Usadel equations\cite{usadel} which are very convenient
when dealing with S/F systems with critical temperatures $T_c$ and exchange
fields $h$ such as $T_c\tau \ll 1$ and $h\tau \ll 1$, where $\tau$ is the
elastic scattering time.

Near the second order transition into superconducting state, Usadel
equations can be linearized with respect to the amplitude of the
superconducting gap. In the S region, the linearized Usadel equation is: 
\begin{equation}  \label{linear-usadel-s}
-D_s\bm{\nabla}^2 \hat{f_s}+2|\omega|\hat{f_s} =2\Delta_s(\bm{r})\hat{\sigma}_z,
\end{equation}
where $D_s$ is the diffusion constant in the superconductor and $\omega=(2
n+1)\pi T$ is a Matsubara frequency at temperature $T$. (Notations are
similar to the ones used in Ref.~\onlinecite{BergEfVolk-rew}, except for the
factor $i$ in front of $\Delta_s$). Eq.~(\ref{linear-usadel-s}) relates the
anomalous Green's function $\hat{f_s}$, which is a matrix in spin space, to
the superconducting gap $\Delta_s(x)$, $\hat{\sigma}_{(x,y,z)}$ are Pauli
matrices in spin space. In the absence of supercurrent, the gap can be taken
as real.

In F region, an exchange field $\bm{h}_{f}=(h_{f,x},h_{f,y},h_{f,z})$ is
acting on the spins of conduction electrons and the linearized Usadel
equation for anomalous function $\hat{f_{f}}$ is 
\begin{eqnarray}
-D_{f}\bm{\nabla}^{2}\hat{f}_{f}+2|\omega |\hat{f}_{f}+is_{\omega }\left(
h_{f,x}[\hat{\sigma}_{x},\hat{f_{f}}]\right.  &&  \nonumber
\label{linear-usadel-f} \\
\left. +h_{f,y}[\hat{\sigma}_{y},\hat{f_{f}}]+h_{f,z}\{\hat{\sigma}_{z},\hat{%
f_{f}}\}\right)  &=&0.
\end{eqnarray}%
Here, $D_{f}$ is the diffusion constant in the ferromagnet and we used the
abbreviation $s_{\omega }\equiv \mathrm{sgn}(\omega )$.

In addition, the boundary conditions at the interfaces with vacuum yield $%
\partial_z \hat{f_s}(z=d_s)=0$ and $\partial_z \hat{f_f}(z=-d_f)=0$, where $%
d_s$ and $d_f$ are the thicknesses of S and F layers, respectively, and $z=0$
defines the plane of the interface between both layers. We also consider
that S and F layers are separated by a thin insulating tunnel barrier.
Therefore, the boundary conditions at the interface $z=0$ are:\cite%
{KL-boundary} 
\begin{equation}  \label{eq:bc}
\sigma_s\partial_z \hat{f_s}(0) = \sigma_f\partial_s \hat{f_f}(0), \quad 
\hat{f_s}(0)=\hat{f_f}(0)+ \gamma_B \xi_s \left.\partial_z \hat{f_f}%
\right|_{z=0},
\end{equation}
where $\sigma_s$ and $\sigma_f$ are conductivities in the layers, and $%
\gamma_B$ is related to the boundary resistance per unit area $R_b$ through $%
\gamma_B \xi_s=R_b \sigma_f$, where $\xi_s$ is the superconducting coherence
length.

The critical temperature $T=T_c$ at the second order transition is now
obtained from the self consistency equation for the gap: 
\begin{equation}  \label{selfconsistent-gap}
\Delta(\bm{r})\ln\frac{T}{T_{c0}} +\pi T \sum_{\omega} \left( \frac{\Delta(%
\bm{r})}{|\omega|}-f_s^{11}(\bm{r},\omega) \right) =0
\end{equation}
($f^{11}$ is a matrix element of $\hat{f}$),
where $T_{c0}$ is the bare transition temperature of the S layer.

In the ferromagnet, the magnitude $h_f$ of the exchange field $\bm{h}_f$ is
fixed. However, its orientation can rotate in the presence of a magnetic
domain wall structure. In the following, we assume a one-dimensional domain
wall structure, along $x$-axis. In order to find the critical temperature of
the bilayer in the presence a domain wall, we must find the $x$- and $z$%
-dependence of the gap and anomalous functions $\hat{f_s}$ and $\hat{f_f}$
which solve Eqs.~(\ref{linear-usadel-s})-(\ref{selfconsistent-gap}).
Proximity effect can significantly affect the transition temperature only
when the thickness of S layer is comparable with the superconducting
coherence length $\xi_s=\sqrt{D_s/2 \pi T_{c0}}$. In order to get tractable
expressions, we will only consider the case $d_s\ll\xi_s$. Such regime is
also well achievable experimentally. Then, $\hat{f}_s$ and $\Delta_s$ are
almost constant along $z$-axis. Therefore, we can average Eq.~(\ref%
{linear-usadel-s}) on the thickness of S layer and make use of the boundary
condition at the interface with vacuum. Finally, we get the following
equation at $z=0$: 
\begin{equation}  \label{eq:usadel-s2}
-D_s\partial^2_x \hat{f_s} +\frac{D_s}{d_s} \partial_z \hat{f_s} +2|\omega|%
\hat{f}_s =2\Delta_s\hat{\sigma}_z.
\end{equation}

The characteristic scale for the proximity effect in F layer is rather set
by coherence length $\xi_f=\sqrt{D_f/h_f}$, where $h_f$ is the typical
amplitude of the exchange field in the ferromagnet. In this section, we will
address the case of very thin F layer: $d_f\ll\xi_f$. Then, from Eq.~(\ref%
{linear-usadel-f}) we can derive similarly the Usadel equation averaged over
the thickness $d_f$, at $z=0$: 
\begin{eqnarray}  \label{linear-usadel-f2}
-D_f \partial^2_x \hat{f}_f -\frac{D_f}{d_f} \partial_z \hat{f_f} +2|\omega|%
\hat{f}_f +i s_\omega \left(h_{f,x}[\hat{\sigma}_x,\hat{f_f}] \right. && \\
\left. +h_{f,y}[\hat{\sigma}_y,\hat{f_f}] +h_{f,z}\{\hat{\sigma}_z,\hat{f_f}%
\} \right)&=&0.  \nonumber
\end{eqnarray}

Let us note right now that the assumption of very thin F layer is quite hard
to achieve experimentally, as we will discuss at the end of Sec. \ref{sec5}.
In Appendix, we will consider the case of arbitrary thickness for the F
layer.

For simplicity, we also consider the case of low interface resistance ($%
\gamma _{B}\rightarrow 0$) where the proximity effect is maximal. In this
regime, $\hat{f}_{f}(x)\approx \hat{f}_{s}(x)\equiv \hat{f}(x)$. By proper
linear combination Eqs.~(\ref{eq:bc}, \ref{eq:usadel-s2}, \ref{linear-usadel-f2}), 
we can form a single equation on $\hat{f}(x)$: 
\begin{eqnarray}
-D\partial _{x}^{2}\hat{f}+2|\omega |\hat{f}+is_{\omega }\left( h_{x}[\hat{%
\sigma}_{x},\hat{f_{f}}]+h_{y}[\hat{\sigma}_{y},\hat{f_{f}}]\right. \qquad  
\nonumber  \label{eq:usadel-eff} \\
\left. +h_{z}\{\hat{\sigma}_{z},\hat{f_{f}}\}\right)  =2\Delta \hat{\sigma}%
_{z},\qquad \Delta =\frac{\eta _{s}}{\eta _{s}+\eta _{f}}\Delta _{s},
\end{eqnarray}%
where $\eta _{s}=\sigma _{s}d_{s}/D_{s}$ and $\eta _{f}=\sigma
_{f}d_{f}/D_{f}$. Therefore, the thin bilayer is described by the same
equations as for a magnetic superconductor,\cite{magsuper} with effective
diffusion constant, exchange field, and BCS coupling constant: 
\begin{equation}
D=\frac{D_{s}\eta _{s}+D_{f}\eta _{f}}{\eta _{s}+\eta _{f}},\quad \bm{h}=%
\frac{\eta _{f}}{\eta _{s}+\eta _{f}}\bm{h}_{f},\quad \tilde{\lambda}=\frac{%
\eta _{s}}{\eta _{s}+\eta _{f}}\lambda ,  \label{renorm}
\end{equation}%
respectively. An equation similar to Eq.~(\ref{eq:usadel-eff}) was derived
for a thin normal-metal/superconductor bilayer, in the absence of exchange
field ($\bm{h}=0$) in Ref.~\onlinecite{fominov}. There, it was shown that
the reduction of the coupling constant $\tilde{\lambda}$ leads to a rapid
decrease of the bilayer critical temperature. In the following, we do not
consider these effects. Rather, we dwell with the case when $\tilde{\lambda}%
\approx \lambda $ and the reduction of $T_{c}$ is mainly due to effective
exchange field $\bm{h}$. Such situation occurs at $\eta _{s}\gg \eta _{f}$.
(When S and F layers have comparable diffusion constant and conductivity,
the renormalization factors in Eq.~(\ref{renorm}) receive a simple
interpretation in terms of volume ratios. In particular, the condition 
$\eta _{s}\gg \eta _{f}$ results in $d_{s}\gg d_{f}$.) Thus, $D\approx D_{s}$ and $\bm{h}%
\approx (\eta _{f}/\eta _{s})\bm{h}_{f}$. Let us note that the amplitude of $%
\bm{h}$ is strongly reduced compared to $\bm{h}_{f}$, eventually it
is of the order of $\Delta _{s}$, and thus it leads to the possible coexistence of
magnetism and superconductivity in the bilayer.

The phase diagram of magnetic superconductors with constant exchange field
was studied long ago.\cite{sarma} Second order transition line from normal
to superconducting state at the critical temperature $T=T_c(h)$ is given by
equation 
\begin{equation}  \label{eq:Tc0}
\ln\frac{T}{T_{c0}}+2\pi T \mathrm{Re} \sum_{\omega>0} \left\{ \frac{1}{%
\omega} -\frac{1}{\omega+ih} \right\}=0,
\end{equation}
where $h$ is the amplitude of $\bm{h}$. At zero temperature, the critical
field is $h_{c}^{(2)}=\Delta_0/2$, where $\Delta_0 \simeq 1.76 T_{c0}$ is
the superconducting gap. However, at $T<T^\ast\approx 0.56 T_{c0}$, the
transition into superconducting state is of the first order and the critical
field at zero temperature is rather $h_c=\Delta_0/\sqrt{2}$.

In the presence of a domain structure in the ferromagnet, the average
exchange field felt by the electrons near domain walls is smaller than in
the domains. This may lead to the enhancement of the superconducting
critical temperature. On the basis of the Usadel equation (\ref%
{eq:usadel-eff}) with self consistency equation (\ref{selfconsistent-gap}),
we consider now this problem in the case of narrow domain walls in Sec. \ref%
{sec3} and large domain walls in Sec. \ref{sec4}.

\section{Narrow domain wall}

\label{sec3}

In this Section, we consider the case of thin domain walls characterized by
the domain wall thickness $w\ll\xi_s$. In Ref.~\onlinecite{BuzBulPan}, the
zero temperature critical field was obtained in the context of magnetic
superconductors. Here, we revise the result and obtain the phase diagram at
finite temperature.

\bigskip

We model the exchange field $\bm{h}$ acting on the electrons with a step
function: $h_{z}(x)=h\,\mathrm{sgn}(x)$, $h_{y}=h_{z}=0$. The structure of
Usadel equation (\ref{eq:usadel-eff}) in spin space simplifies greatly and 
we have: 
\begin{equation}
-\frac{D}{2}\partial _{x}^{2}f^{11}+\left( |\omega |+is_{\omega
}h_{z}(x)\right) f^{11}=\Delta ,  \label{eq:usadel-thin}
\end{equation}%
while $f^{12}=f^{21}=0$ and $%
f^{22}_\omega=-f^{11}_{-\omega}$. Its solution for a given $\Delta $ is 
\begin{equation}
f^{11}(x)=\int dy\mathcal{G}(x,y)\Delta (y),  \label{sol-f}
\end{equation}%
where $\mathcal{G}$ is the Green's function associated with the homogeneous
differential equation (\ref{eq:usadel-thin}) ; $\mathcal{G}$ is defined by: 
\begin{eqnarray}
\mathcal{G}(x,y) &=&\frac{e^{-\kappa x}}{D\kappa }\left[ e^{\kappa y}+\frac{%
\kappa -\kappa ^{\ast }}{\kappa +\kappa ^{\ast }}e^{-\kappa y}\right] \quad 
\text{for}\quad x>y>0,  \nonumber \\
\mathcal{G}(x,y) &=&\frac{2}{D(\kappa +\kappa ^{\ast })}e^{-\kappa
x}e^{\kappa ^{\ast }y}\quad \text{for}\quad x>0>y,
\end{eqnarray}%
where $\kappa =\sqrt{2(|\omega |+is_{\omega }h)/D}$, while $\mathcal{G}(x,y)=%
\mathcal{G}(y,x)$ and $\mathcal{G}(x,y)=\mathcal{G}(-x,-y)^{\ast }$.

We look for a symmetric solution $\Delta(-x)=\Delta(x)$. Writing the self
consistency equation (\ref{sol-f}) in Fourier space, we get the equation
defining the critical temperature $T=T_{cw}$ for DWS formation: 
\begin{eqnarray}  \label{eq:gap-thin2}
\left( \ln\frac{T}{T_{c0}} +2\pi T \mathrm{Re} \sum_{\omega>0} \frac{1}{%
\omega} -\frac{1}{\omega+ih+\frac{D p^2}{2}} \right)\Delta_p= \qquad \\
2T \sum_{\omega>0}\int dk \frac{h\sqrt{D}\left[(\omega^2+h^2)(\sqrt{%
\omega^2+h^2}-\omega)\right]^{\frac{1}{2}}} {[h^2+(\omega+\frac{D p^2}{2}%
)^2][h^2+(\omega+\frac{D k^2}{2})^2]} \Delta_k.  \nonumber
\end{eqnarray}

Close to $T_{c0}$, at $h\ll T_{c0}$, the critical temperature for the
transition into uniform superconducting state can be obtained analytically
from Eq.~(\ref{eq:Tc0}): 
\begin{equation}  \label{eq:hc}
\frac{T_{c0}-T_c(h)}{T_{c0}}= \frac{7\zeta(3)}{4\pi^2}\frac{h^2}{T_{c0}^2} -%
\frac{31\zeta(5)}{16\pi^4}\frac{h^4}{T_{c0}^4}+\dots
\end{equation}
On the other hand, Eq.~(\ref{eq:gap-thin2}) for the DWS can be simplified: 
\begin{equation}
\left( \frac{T_{cw}(h)-T_c(h)}{T_{c0}} +\frac{\pi}{8} \frac{D p^2}{T_{c0}}
\right) \Delta_p= A \frac{h^2}{T_{c0}^2} \sqrt{\frac{D}{\pi T_{c0}}} \int dk
\Delta_k,
\end{equation}
where $A=(8\sqrt{2}-1)\zeta(\frac{7}{2})/(8\pi^3)$. Such equation is solved
straightforwardly and we get the increase of critical temperature near $%
T_{c0}$ due to DWS: 
\begin{equation}
\frac{T_{cw}(h)-T_c(h)}{T_{c0}} \approx 8A^2\frac{h^4}{T_{c0}^4}.
\end{equation}
The corresponding shape of the order parameter near the transition is given
by 
\begin{equation}
\Delta(x) \sim \exp \left[ -B \frac{|x|}{\xi_s} \frac{T_{c0}-T_{cw}(h)}{%
T_{c0}} \right],
\end{equation}
where $B=16\pi \sqrt{2} A/(7\zeta(3))$. Thus, near $T_{c0}$, the localized
superconductivity is characterized by exponential decay without oscillation
of the superconducting order parameter, with its maximum at the domain wall
position.

\begin{figure}[tbp]
\includegraphics[scale=1]{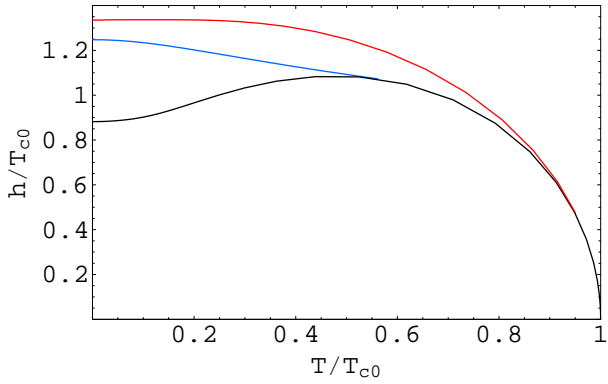}
\caption{Phase diagram. The critical line $h_c^{(2)}(T)$ at the second order
transition into uniform superconducting state is plotted in black. At $%
T<T^*=0.56T_{c0}$, transition into uniform state is of the first order, with
the critical line $h_c(T)$ plotted in blue. The critical line $h_{cw}(T)$
corresponding to domain wall superconductivity is of the second order and is
plotted in red. }
\label{fig:1}
\end{figure}

Away from $T_{c0}$, Eq.~(\ref{eq:gap-thin2}) does not contain any small
parameter. Its structure is that of a linear integral equation whose kernel
is a superposition of separable terms. Such form of the kernel is known to
be convenient for numerical calculation. As a result, we obtained the second
order critical line at any temperature (cf. Fig.~\ref{fig:1}). The critical
line is significantly increased compared to the critical line for the
transition into uniform superconducting state. In particular, at zero
temperature, the critical field for localized superconductivity at $T=0$ is $%
h_{cw}\simeq 1.33T_{c0}\sim 0.76\Delta _{0}$ and lies above the critical
field for the first order transition into uniform state, $h_{c}\simeq
0.71\Delta _{0}$. It is of interest to note that, if the effective exchange
field in the bilayer is between $h_{c}$ and $h_{cw}$, then the special
situation occurs when only DWS can be realized in the system, but no
superconductivity faraway from the domain walls.

\begin{figure}[tbp]
\includegraphics{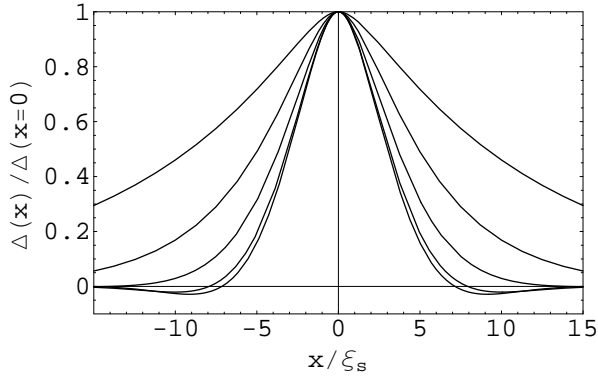}
\caption{Self consistent gap just below the second order transition line $%
h_{cw}(T)$ into localized domain wall superconducting state at different
temperatures $T=0,0.2,0.4,0.6,0.8\,T_{c0}$ (from narrowest to widest).}
\label{fig:2}
\end{figure}

We also plot the self consistent order parameter at different temperatures
in Fig.~\ref{fig:2}. In addition to the decay, it shows small oscillations
along the direction perpendicular to the domain wall at low temperatures.

As it was pointed out in Sec. \ref{sec2}, transition into uniform
superconducting state happens to be of the first order at $T<T^{\ast }$ in
the bilayer. Therefore, one should also worry about the possibility for the
change of the transition order along the critical line corresponding to DWS.
We considered this possibility by solving the nonlinear Usadel equations
perturbatively up to the third order terms in the gap $\Delta $. We found that
such equation corresponds to the saddle-point of the free energy density
(per unit thickness of the bilayer and per unit length of the magnetic
domain wall) functional: 
\begin{eqnarray}
\mathcal{F} &=&\mathcal{F}_{2}+\mathcal{F}_{4}+\dots   \label{eq:free-energy}
\\
\mathcal{F}_{2} &=&\nu \int dx\left\{ |\Delta |^{2}\ln \frac{T}{T_{c0}}+2\pi
T\Re \sum_{\omega >0}\frac{|\Delta |^{2}}{\omega }-\Delta ^{\ast }f_{\omega
}^{11}\right\}   \nonumber \\
\mathcal{F}_{4} &=&\frac{\pi T\nu }{2}\Re \sum_{\omega >0}\int dx\left[ 
\frac{D}{2}(\partial _{x}f^{11}_\omega )^{2}+\Delta f^{11}_\omega \right] (\bar{%
f}^{11}_\omega)^2 ,  \nonumber
\end{eqnarray}%
where $\nu $ is the density of states at the Fermi level in the normal
state, and $\bar{f}_{\omega }^{11}=(f_{-\omega }^{11})^{\ast }$ where $%
f_{\omega }^{11}$ is given by Eq.~(\ref{sol-f}). At the second order
transition, the term $\mathcal{F}_{2}$, which is quadratic in $\Delta $,
vanishes when Eq.~(\ref{eq:gap-thin2}) is satisfied. The phase transition is
stable provided that the term $\mathcal{F}_{4}$, which is quartic in $\Delta 
$, remains positive along the transition line. We checked numerically that
this was indeed the case. In particular, at $T=0$ and $h$ close to $h_{cw}$,
we found: 
\begin{eqnarray}
\mathcal{F}_{2} &=&\nu \ln \frac{h}{h_{cw}}\int dx|\Delta (x)|^{2}  \nonumber
\\
&\simeq &2.87\nu \sqrt{\frac{D}{2h_{cw}}}\ln \frac{h}{h_{cw}}\Delta
(x=0)^{2},  \nonumber \\
\mathcal{F}_{4} &=&0.26\nu \sqrt{\frac{D}{2h_{cw}}}\frac{\Delta (x=0)^{4}}{%
h_{cw}^{2}}.
\end{eqnarray}

Thus we found that the transition into DWS remains of the second order at
all temperatures.

\section{Large domain wall}

\label{sec4}

In this section, we consider the case of a large domain wall with thickness $%
w\gg \xi_s$. The domain wall can be described with an exchange field $\bm{h}%
=h(\cos \phi,\sin \phi,0)$, the rotation angle $\phi(x)$ varies monotonously
between $\phi(-\infty)=-\pi/2$ and $\phi(\infty)=\pi/2$. We find that the
critical temperature at DWS nucleation is given by a Schr\"odinger equation
for a particle in the presence of a potential well whose profile is
proportional to $-(\partial_x\phi)^2$. This result is not specific to the
thin bilayer considered in this Section. In the Appendix, we extend it to
bilayers with arbitrary thickness of the F layer and arbitrary transparency
of the S/F interface.

\bigskip

At the transition, the linearized Usadel equation (\ref{eq:usadel-eff}) must
be solved, that is: 
\begin{eqnarray}  \label{usadel-large}
&&-\frac{D}{2}\partial_x^2 f^{11}+|\omega|f^{11} -i\frac{h s_\omega}{2}
\left( e^{-i\phi}f^{21} -e^{-i\phi}f^{12}\right) =\Delta  \nonumber \\
&&-\frac{D}{2}\partial_x^2 f^{12}+|\omega| f^{12} +ihs_\omega
e^{-i\phi}f^{11} =0  \nonumber \\
&&-\frac{D}{2}\partial_x^2f^{21}+|\omega|f^{21} -ihs_\omega e^{i\phi}f^{11}
=0.
\end{eqnarray}
while $f^{22}=-f^{11}$.

If the spatial dependence of $\phi $ is neglected, the gap $\Delta $ is
uniform along the layer and the solutions are readily found: 
\begin{equation}
f_{0}^{11}=\frac{\Delta |\omega |}{\omega ^{2}+h^{2}},\quad f_{0}^{12}=\frac{%
i\Delta hs_{\omega }e^{-i\phi }}{\omega ^{2}+h^{2}}=-e^{-2i\phi }f_{0}^{21},
\label{eq:ordre0}
\end{equation}%
As a result, the critical temperature $T_{c}$ into uniform superconducting
state naturally does not depend on $\phi $ and is again given by
Eq.~(\ref{eq:Tc0}).

Now, let us assume that $\phi$ varies slowly. We solve Usadel equation (\ref%
{usadel-large}) perturbatively by looking for a solution $\hat{f}\approx\hat{%
f}_0+\hat{f}_1$. In first approximation, $\hat{f}_0$ is still given by Eq.~(%
\ref{eq:ordre0}), where $\phi$, and, possibly, $\Delta$, now slowly depend
on $x$. The correction $\hat{f}_1$ induced by their spatial dependence 
is determined by the set of equations: 
\begin{eqnarray}
&&|\omega|f^{11}_1 -i\frac{h s_\omega}{2} \left( e^{-i\phi}f^{21}_1
-e^{-i\phi}f^{12}_1\right) =\frac{D}{2}\partial_x^2 f^{11}_0  \nonumber \\
&&|\omega| f^{12}_1 +ihs_\omega e^{-i\phi}f^{11}_1 =\frac{D}{2}\partial_x^2
f^{12}_0  \nonumber \\
&& |\omega|f^{21}_1 -ihs_\omega e^{i\phi}f^{11}_1 =\frac{D}{2}\partial_x^2
f^{21}_0,
\end{eqnarray}
By appropriate linear combination of these equations, one finds that 
\begin{equation}
f^{11} \approx \frac{|\omega|}{\omega^2+h^2} \Delta + \frac{Dh^2}{%
2(\omega^2+h^2)^2}(\phi^{\prime})^2 \Delta + \frac{D}{2} \frac{\omega^2-h^2}{%
(\omega^2+h^2)^2} \Delta^{\prime\prime}.
\end{equation}
Here, primes stand for derivative along $x$. Inserting this solution in the
self consistency equation (\ref{selfconsistent-gap}) we obtain the equation
for the gap 
\begin{equation}  \label{eq:schrodinger}
-\frac{1}{2m}\Delta^{\prime\prime}(x)+U(x)\Delta(x)=E\Delta(x),
\end{equation}
where 
\begin{eqnarray}  \label{coef-schrod}
E=-\ln\frac{T}{T_c}, \quad \frac{1}{2m}=D\pi T \sum_{\omega>0}\frac{%
(\omega^2-h^2)}{(\omega^2+h^2)^2},  \nonumber \\
U(x)=-D\pi T(\phi^{\prime})^2 \sum_{\omega>0}\frac{h^2}{(\omega^2+h^2)^2}.
\end{eqnarray}

Equation (\ref{eq:schrodinger}) is a linearized Ginzburg-Landau equation for
a magnetic superconductor in the presence of a domain wall. It can be easily
checked that effective mass $m$ is always positive and $U(x)$ is negative.
Therefore, Eq.~(\ref{eq:schrodinger}) looks like a one-dimensional
Schr\"odinger equation for a particle in potential well $U(x)$. It is well
known that a bound state with $E<0$ always forms in such potential. As a
result, second order transition into localized superconducting state always
appears more favorable than into uniform state.

Let us estimate now the magnitude of critical temperature increase.

Close to $T_{c0}$, the spatial variation for $\Delta$ is set by the
temperature-dependent coherence length $\xi(T)=\xi_s \sqrt{T_{c0}/(T_{c0}-T)}
$ which diverges at the transition. Therefore, $\xi_s\ll w\ll\xi(T)$ and the
potential well can be approximated by a delta-potential: $U(x)\approx -(\pi
D h^2 /48 T_{c0}^3 w)\delta(x)$, while $1/2m\sim(\pi D /8 T_{c0})$.
Therefore we get the estimate: 
\begin{equation}  \label{eq:result}
\frac{T_{cw}-T_{c}}{T_c} 
\simeq 
\frac{\pi^6}{36} 
\left(\frac{h}{2 \pi T_{c0}}\right)^4
\frac{\xi_s^2}{w^2} 
\propto 
\frac{\xi_s^2}{w^2} 
\frac{(T_{c0}-T_c)^2}{T_{c0}^2}.
\end{equation}
Such increase is small in the ratio $(\xi_s/w)^2$, and another reduction
factor comes from the smallness of critical exchange field $h$ near $T_{c0}$%
, see Eq.~(\ref{eq:hc}).

At $T\rightarrow 0$, the second order transition into uniform state occurs
at exchange field $h_{c}^{(2)}=\Delta _{0}/2$. On the other hand, the
effective mass in Eq.~(\ref{eq:schrodinger}) diverges (as $1/2m\sim (\pi
D/T)e^{-h/T}$). The fact that $m$ remains positive at finite temperatures is
related to the absence of instability toward a modulated superconducting
(Fulde-Ferrell-Larkin-Ovchinnikov) state in magnetic superconductors in the
presence of strong disorder, $\tau T_{c0}\ll 1$.\cite{aslamazov} Due to its
large inertia, the particle now resides in the minimum of the potential well
(zero point fluctuations can be neglected): $U_{\text{min}}=-(\pi D/16hw^{2})
$. The corresponding increase in critical exchange field at $T=0$ is $%
(h_{cw}-h_{c}^{(2)})=\pi D/16w^{2}$. Such increase is of the order of
magnitude $\sim (\xi _{s}/w)^{2}h_{c}^{(2)}\ll h_{c}^{(2)}$. Actually, at 
low temperature the transition into the superconducting state in
the uniform exchange field is a first order. We may expect that the domain
wall superconductivity in such situation also appears by a first order
transition.

At intermediate temperature, we may obtain the critical temperature $%
T=T_{cw} $ with a specific choice of the spatial dependence of rotation
angle $\phi$. Assuming that $\phi(x)=2 \arctan [\tanh (\xi/2w)]$, we get:%
\cite{landau3} 
\begin{equation}
\ln\frac{T}{T_c} = \frac{T_{c0}}{8 T_c} \frac{\xi_s^2}{w^2} F(\frac{h}{%
2\pi T_c}),
\end{equation}
where 
\begin{equation}
F(u)=\Re\psi_1(Z) \left[-1+\sqrt{1-\frac{2 \Re (i \psi(Z)+u\psi_1(Z))}{%
u\Re\psi_1(Z)}}\right]^2,
\end{equation}
where $\psi$ and $\psi_1$ are digamma functions, and $%
Z=1/2+i u$.

We should note however that transition into uniform superconducting state
becomes of the first order at $T<T^*$ and results in significant increase of
the critical line $h_c(T)$. Most probably, such change of the transition
order should also be considered for localized superconductivity.

\section{Discussion}

\label{sec5}

The qualitative picture of the effect of domains walls on the
superconducting properties now emerges from our calculations.

First, the electrons of the Cooper pairs which travel across the S/F
interface feel the exchange field $h_{f}$ in the F layer. This proximity
effect results in an effective pair breaking that weakens superconductivity
in S layer. As a result, the critical temperature $T_c$ of the bilayer gets
suppressed in comparison with the critical 
temperature $T_{c0}$ of the bare S film: $T_{c0}-T_c\sim\tau _{s}^{-1}$. 
There, the pair-breaking time $\tau_s$ can be estimated from Eq.~(\ref{eq:hc}): at $h\ll T_{c0}$, 
$\tau _{s}^{-1}\sim h^{2}/T_{c0}$, where  $h$ is the effective
exchange field that would act directly in the bare S layer to yield the same
pair breaking effect due to $h_{f}$ in the bilayer. 
In particular, when the S/F interface
is transparent and F layer is thin, we found that $h\sim
h_{f}d_{f}/(d_{s}+d_{f})$, where $d_{f}$ and $d_{s}$ are the thicknesses of
F and S layers, respectively. On the other hand, when $h\gg T_{c0}$, there is 
no superconducting transition in the bilayer.

Second, in the vicinity of magnetic domain walls in the F layer, such pair
breaking mechanism becomes less effective. Therefore, localized
superconductivity may appear with critical temperature $T_{cw}>T_{c}$. 
When the domain wall width $w$ is large in comparison with supercondutcing 
coherence length $\xi_s$, the exchange field rotates by the angle $\theta \sim
\xi_s/w$ on the scale of proximity effect. Therefore, the decrease of the 
average exchange field close to the wall is estimated as $h-h_{av} \sim \theta^2 h$. 
Correspondingly, the pair-breaking time is increased by 
$\Delta \tau_s/\tau_s \sim \theta^2$. In
analogy with the theory of superconductivity at twinning-plane boundaries%
\cite{TP}, one can estimate the increase of $T_{c}$:
\begin{equation}
\Delta T
\equiv
T_{cw}-T_{c} 
\sim
\frac{\theta^2}{\tau_s}
\frac{w}{\xi(T)}.
\end{equation}
Here, the temperature dependent correlation length $\xi (T)\sim \xi _{s}\sqrt{%
T_{c}/\Delta T}$ is the spatial extension of the superconducting gap and it
diverges close to the superconducting transition; the pair breaking is only
reduced on the small portion of the gap corresponding to the width $w$
of the domain wall. On the end, the formula yields the estimate $\Delta T/T_{c}\sim (\xi
_{s}/w)^{2}/(\tau _{s}T_{c})^{2}$. Combining this result with the estimates
for $\tau _{s}$ given in the preceding paragraph, we finally retrieve Eq.~(%
\ref{eq:result}) qualitatively. This result holds when the width $w$ is much
larger than $\xi _{s}$. At smaller domain wall width, weakening of pair breaking 
effect works on the characteristic scale $\xi _{s}$ of the proximity effect 
and $w$ should be replaced by $\xi _{s}$
in the estimate. Therefore, the large enhancement of $T_{cw}$, of
the order of $T_{c})$, can be expected when $w\lesssim \xi _{s}$ and $%
h\sim T_{c0}$.

In the present work, we confirmed quantitatively this estimate in a
number of situations. In particular, we obtained that, in the case of strong
enhancement of $T_{cw}$ (at $w\lesssim \xi _{s}$), for a thin F layer,
transition into DWS state remains of the second order, with critical line
above the transition into uniform superconducting state of the second order
at $T>T^{\ast }=0.56T_{c0}$, and of the first order at $T<T^{\ast }$. We
predicted that only DWS could appear in bilayers with appropriate parameters.

\bigskip

Let us now come back to the assumption of very thin F layer that was taken
in the calculations. Usually, the exchange field in a ferromagnet is much
larger than the gap in a superconductor. Therefore, the coherence length $%
\xi_f$ is much smaller than $\xi_s$ ($1-5\,\text{nm}$ for the former, compared to $%
10-50 \,\text{nm}$ for the latter). Therefore, the regime $d_f\ll\xi_f$ is quite
hard to achieve with real samples of diffusive ferromagnets. Extending the
calculation for thin F layer to arbitrary thickness of the layer and
arbitrary transparency (characterized by $\gamma_B$) of the interface is quite 
straightforward when domain walls are large, as we show in the
Appendix. As it is well known, the behavior of the critical temperature of
the transition into uniform superconducting state is quite rich in such case
and may even oscillate as a function of the parameters such as $d_f$ or $%
\gamma_B$. However, the physics of DWS appears to be quite similar to the
one derived for thin bilayer. In particular, in the case of thick F layer ($%
d_f\gg\xi_f=\sqrt{D_f/h_f}$), the effective exchange field that would enter
the above qualitative estimates would be $h\sim(\xi_s/d_s)\sqrt{h_f T_{c0}}$%
. Clearly, at $\xi_s \gg d_s$, as we assumed from the beginning, such field
is much larger than $T_{c0}$ and leads therefore to superconductivity
suppression. However, $h$ is strongly reduced if the S/F interface is
opaque, which may lead to the F/S coexistence and to DWS appearance, as
studied in this work. Nevertheless, the enhancement of $T_c$ still is small
by the factor $\xi_s/w\ll 1$ in the situation described in the Appendix.

These considerations suggest two possible directions to extend the range of
existence of DWS. In the present work, DWS was analyzed for S layers with 
thickness $d_s$ much smaller than coherence length $\xi_s$. On the other hand,
DWS should not appear when the superconductor is hardly affected by proximity 
effect, at $d_s \gg \xi_s$. It would be of interest to consider the intermediate
case when $d_s$ and $\xi_s$ are of the same order. This was
studied for instance in the absence of magnetic domains in Ref.~%
\onlinecite{fominov2}. Maybe a more important point would be to address the
case of DWS in S/F bilayer with narrow domain wall and large thickness of
the F layer. However, both problems require considerably more numerical
work, which goes beyond the scope of this paper.

\bigskip

A lot of attention has been devoted to the study of long range triplet
proximity effect which develops in S/F structures when the direction of
exchange field in the ferromagnet varies spatially\cite%
{BergEfVolk-rew,champel,volkov}. In our calculation, such long range triplet
component is also present, as it is clear from the Appendix: in Eq.~(\ref%
{eq:sol-gen}), the term $\gamma_0\mathrm{ch}q_0\tilde{z}$ in the direction
transverse to the bilayer is generated only because of the presence of
domain wall, and it decays with typical length $\xi_T\sim\sqrt{D_f/T}\gg
\xi_f$. However, $\gamma_0$ does not enter $f^{11}$ and, therefore, is not
important for the determination of the critical temperature of transition
into DWS. In all the calculations we presented, there is no long range
triplet component in the direction transverse to the wall either: the
typical length for the superconducting gap is determined by conventional
short range proximity effect with decay length $\sim \xi_s$. We would like
to emphasize also that the $T_c$ enhancement due to the appearance of DWS is
maximized for narrow domain wall (see Sec. \ref{sec3}), when the matrix
elements of the anomalous Green's function that would give rise to long
range triplet component are exactly zero. Therefore, the physics of DWS
discussed here is not directly related to such phenomenon.

This observation is in agreement with Ref.~\onlinecite{champel} where the
calculation of critical temperature of S/F bilayers in the presence of
spiral magnetic order in the F layer was presented. There also, long range triplet
component was found not to contribute to the result. On the other hand, long
range triplet component may be important for other properties such as
density of state in the ferromagnetic layer with domain structure deposited
on top of a bulk superconducting substrate.\cite{volkov}

We would like also to emphasize that our calculations differ from the study
of S/F bilayers in the presence of spiral magnetic order in the F layer.\cite%
{champel} These works can be interpreted as considerations on magnetic
domain structure only in as so much that the width of the domains $L$ and
the width of the walls $w$ are identical. A consequence is that the
superconducting gap is spatially uniform along the bilayer in Ref.~%
\onlinecite{champel}. In contrast, our theory really shows that, in the more
realistic case when $w\ll L$, truly localized superconducting states can
appear. In addition, consideration on the effect of spiral magnetic order
corresponding to $\phi(x)=Q x$ in Eqs.~(\ref{eq:schrodinger},\ref%
{coef-schrod}), where $Q$ is the wavevector of the spiral, can be
immediately calculated from the Schr\"odinger-like equation (\ref%
{eq:schrodinger}), at least when $Q\xi_s\ll 1$. Critical temperature
enhancement due to spiral magnetic order and corresponding to uniform gap $%
\Delta(x)$ follows straightforwardly from the observation that $%
\phi^{\prime}(x)=Q$ which enters Eq.~(\ref{coef-schrod}) is constant.

\bigskip

In conclusion, we analyzed in this work the enhancement
of the superconducting critical temperature of superconducting/ferromagnetic
bilayers due to the appearance of localized superconducting states in the
vicinity of magnetic domain walls in the ferromagnet. We considered
the case when the main mechanism of the superconductivity destruction via
the proximity effect is the exchange field. We demonstrated that the
influence of the domain walls on the superconducting properties of S layer
may be quite strong if the domain wall thickness is of the order of
superconducting coherence length.

We interpreted qualitatively and quantitatively the amplitude of this effect, 
and we pointed out the special case when parameters of the bilayer are
such that only localized superconductivity may form in these systems.

For magnetic film with perpendicular anisotropy, the orbital effect
provides an additional mechanism for the domain wall superconductivity\cite{BuzMel}
and it may be easily taken into account. On the other hand for the film 
with the easy plane magnetic anisotropy the domain wall will be a source 
of the magnetic field in the
adjacent S layer and locally weakens the superconductivity.\cite{sonin}  
The role of the orbital mechanism in the domain wall superconductivity may be important only
if the magnetic induction is comparable with the upper critical field of the
superconducting film.

The domain wall superconductivity in S/F bilayers opens an
interesting way to manipulate the superconducting properties through the
domain structure. In particular the motion of the domain wall in F layer may
be accompanied by the displacement of the narrow superconducting region in S
layer.

\bigskip

We are grateful to J. Aarts for attracting our attention to the
problem of the domain wall superconductivity in S/F bilayers and useful
discussions.

\appendix

\section{Thick F layer}

\label{app}

As mentioned previously, $h_f$ is usually large in ferromagnets. Therefore
the condition of thin F layer, $d_f \ll \xi_f$, is hardly reached. We would
like to extend the results of the previous section to the more realistic
situation of a finite size F layer. The difficulty is that the set of
differential equations (\ref{linear-usadel-s})-(\ref{selfconsistent-gap}) to
be solved are now two-dimensional. We managed to solve it for the case of
large domain wall only. In the F layer, we parametrize the exchange field
rotation with a slowly varying angle $\phi(x)$ such as $\bm{h}%
_f=h_f(\cos\phi,\sin\phi,0)$. Linearized Usadel equations (\ref%
{linear-usadel-f}) in F layer are: 
\begin{eqnarray}
&&-\frac{D_f}{2}\triangle f^{11}+|\omega|f^{11} -i\frac{h_f s_\omega}{2}
\left( e^{-i\phi}f^{21} -e^{-i\phi}f^{12}\right) =0,  \nonumber \\
&&-\frac{D_f}{2}\triangle f^{12}+|\omega| f^{12} +ih_fs_\omega
e^{-i\phi}f^{11} =0,  \nonumber \\
&&-\frac{D_f}{2}\triangle f^{21}+|\omega|f^{21} -ih_fs_\omega
e^{i\phi}f^{11} =0.
\end{eqnarray}
where $\triangle=\partial_x^2+\partial_z^2$, while $f^{22}=-f^{11}$. When
the spatial dependence of $\phi$ is neglected, the general form of the
solutions which satisfy boundary condition at the F/vacuum interface are: 
\begin{eqnarray}  \label{eq:eigen}
f^{11}_0&=&F_+\mathrm{ch} q_+\tilde{z}+F_-\mathrm{ch} q_-\tilde{z}, \\
f^{12}_0&=&s_\omega e^{-i\phi} \left\{ F_0\mathrm{ch}q_0\tilde{z}+ F_+%
\mathrm{ch}q_+\tilde{z}-F_-\mathrm{ch}q_-\tilde{z} \right\},  \nonumber \\
f^{21}_0&=&-s_\omega e^{i\phi} \left\{ -F_0\mathrm{ch}q_0\tilde{z}+ F_+%
\mathrm{ch} q_+\tilde{z}-F_-\mathrm{ch} q_-\tilde{z} \right\},  \nonumber
\end{eqnarray}
where $\tilde{z}=z+d_f$, $q_0=\sqrt{2|\omega|/D_f}$, and $q_\pm=\sqrt{2(\pm
ih_f+|\omega|)/D_f}$.

We determine now the amplitudes of the eigenmodes $F_0$ and $F_\pm$. For
this, we first determine $f^{11}$ and $\partial_z f^{11}$ at $z=0$ in F
layer. Making use of the boundary conditions (\ref{eq:bc}), we can now
insert them in the Usadel equation (\ref{eq:usadel-s2}) in the S layer. We
proceed similarly for $f^{12}$ and $f^{21}$. On the end, we get from (\ref%
{eq:usadel-s2}) three equations which determine the amplitudes we are
looking for. We find $F_0=0$, while 
\begin{equation}
F_\pm=\frac{\Delta}{2\Omega_\pm}, \qquad \Omega_\pm=|\omega|C_\pm+\alpha
q_\pm\mathrm{sh}q_\pm d_f,  \label{eq:eigenamplitudes}
\end{equation}
where $C_\pm=\mathrm{ch}q_\pm d_f+\gamma_B\xi_sq_\pm\mathrm{sh}q_\pm d_f$
and $\alpha=D_s\sigma_f/2d_s\sigma_s$.

Inserting Eqs. (\ref{eq:eigen})-(\ref{eq:eigenamplitudes}) into (\ref{eq:bc}%
) to determine $\hat{f}_s$, and then inserting $f^{11}_s$ in the self
consistency equation (\ref{selfconsistent-gap}), we get the equation
defining the critical temperature $T=T_c(h)$ for uniform superconducting
state: 
\begin{eqnarray}
0 &=& \ln\frac{T}{T_{c0}}+2 \pi T \mathrm{Re} \sum_{\omega>0} \left\{ \frac{1%
}{|\omega|} - \frac{1}{|\omega|+\Gamma_+} \right\},  \nonumber \\
\Gamma_+ &=& \frac{\alpha q_+}{\coth q_+ d_f+\gamma_B \xi_s q_+} .
\label{eq:Tc-thick}
\end{eqnarray}
This result is the same as Eq.~(47) of Ref.~\onlinecite{Buzdin-rew}. Whether
transition is of the second order (as described by Eq.~(\ref{eq:Tc-thick}))
or of the first order was considered in Ref.~\onlinecite{tollis}.

Let us now consider the effect of a domain wall. As in section \ref{sec4}, we
will look for a solution $\hat{f}\approx\hat{f}_0+\hat{f}_1$. In leading
order, the spatial dependence of $\phi$ and the gap $\Delta$ is ignored, and 
$\hat{f}_0$ is still given by Eqs.~(\ref{eq:eigen},\ref{eq:eigenamplitudes}%
). The correction $\hat{f}_1$ accounts for the slow variations of $\phi$ and 
$\Delta$ along $x$-axis; it is determined by the set of equations: 
\begin{eqnarray}
-\frac{D_f}{2}\partial_z^2 f^{11}_1+|\omega|f^{11}_1 -i\frac{h_f s_\omega}{2}
\left( e^{-i\phi}f^{21}_1 \right. && \\
\left. -e^{-i\phi}f^{12}_1\right) &=&\frac{D_f}{2}\partial_x^2 f^{11}_0, 
\nonumber \\
-\frac{D_f}{2}\partial_z^2 f^{12}_1+|\omega| f^{12}_1 +ih_fs_\omega
e^{-i\phi}f^{11}_1 &=&\frac{D_f}{2}\partial_x^2 f^{12}_0,  \nonumber \\
-\frac{D_f}{2}\partial_z^2 f^{21}_1+|\omega|f^{21}_1 -ih_fs_\omega
e^{i\phi}f^{11}_1 &=&\frac{D_f}{2}\partial_x^2 f^{21}_0.  \nonumber
\end{eqnarray}
Ignoring $x$-dependence of the right-hand-side of the above differential
equations, we can find the exact $z$-dependent function $\hat{f}_1$ which
solves them. We obtain: 
\begin{eqnarray}  \label{eq:sol-gen}
f^{11} &=& \sum_{a=\pm} \left[ \mathrm{ch} q_a\tilde{z} \left( \frac{\Delta}{%
2\Omega_a} +a\frac{iD_f\phi^{\prime 2}\Delta}{16h_f\Omega_a}+\gamma_a \right)
\right.  \nonumber \\
&& \qquad \qquad \left. -\tilde{z} \mathrm{sh} q_a\tilde{z} \frac{%
\Delta^{\prime\prime}-\frac{1}{2}\phi^{\prime 2}\Delta} {4\Omega_a q_a} %
\right],  \nonumber \\
f^{12,21}&=&\pm s_\omega e^{\mp i\phi} \left\{ \pm \gamma_0 \mathrm{ch} q_0%
\tilde{z} +\sum_{a=\pm} \left[ \left( \frac{a\Delta}{2\Omega_a}+a\gamma_a
\right.\right.\right.  \nonumber \\
&&\left.\left.\left. -\frac{iD_f\phi^{\prime 2}\Delta}{16h_f\Omega_a} \pm 
\frac{D_f(2\phi^\prime\Delta^\prime+\phi^{\prime\prime}\Delta)}{4h_f\Omega_a}
\right) \mathrm{ch} q_a\tilde{z} \right.\right.  \nonumber \\
&& \left.\left. -a \tilde{z} \mathrm{sh} q_a\tilde{z} \frac{%
\Delta^{\prime\prime}-\frac{1}{2}\phi^{\prime 2}\Delta} {4\Omega_a q_a} %
\right]\right\},
\end{eqnarray}
where the integration constants $\gamma_0$ and $\gamma_\pm$ still remain to
be determined. For this purpose, we insert Eq.~(\ref{eq:sol-gen}) into (\ref%
{eq:bc}) in order to get $\hat{f}_s$ and $\partial_z\hat{f}_s$. Inserting
them in Eq.~(\ref{eq:usadel-s2}), we thus obtain a set of three equations
which allow to determine them. In particular, we find: 
\begin{eqnarray}  \label{integ-cst}
\gamma_\pm &=& \pm \frac{iD_f\phi^{\prime 2}\Delta}{16h_f\Omega_\pm} + \frac{%
\Xi_\pm}{4 q_\pm \Omega_\pm^2}(\Delta^{\prime\prime}-\frac{1}{2}\phi^{\prime
2}\Delta)  \nonumber \\
&& +\frac{D_s}{4} \left( \frac{C_\pm}{\Omega_\pm^2} \Delta^{\prime\prime}
\mp \frac{i\phi^{\prime 2} \Delta}{\Omega_\pm} \Im \frac{C_\pm}{\Omega_\pm}
\right)
\end{eqnarray}
where $\Xi_\pm = |\omega | S_\pm + \alpha ( \mathrm{sh} q_\pm d_f + d_f
q_\pm \mathrm{ch} q_\pm d_f)$ and $S_\pm = d_f \mathrm{sh} q_\pm d_f +
\gamma_B \xi_s ( \mathrm{sh} q_\pm d_f + d_f q_\pm \mathrm{ch} q_\pm d_f)$.

Finally, we can insert Eqs.~(\ref{integ-cst}) into (\ref{eq:sol-gen}), and
then into (\ref{eq:bc}), in order to obtain $f^{11}$ in the S alyer. Then, we 
insert it the self consistency equation (\ref{selfconsistent-gap}). On the
end, we find that the superconducting gap at the transition into
DWS state is still determined by the Schr\"{o}dinger equation (\ref%
{eq:schrodinger}), with effective coefficients: 
\begin{eqnarray}
E &=&-\ln \frac{T}{T_{c}},  \label{coef:thick} \\
\frac{1}{2m} &=&\pi T\Re \sum_{\omega >0}
\left(
\frac{D_{s}}{\omega _{+}^{2}}
+
\frac{\alpha (\mathrm{sh}2q_{+}d_{f}+2q_{+}d_{f})}{2q_{+}\Omega_{+}^{2}}%
\right)   \nonumber \\
U(x) &=&-\pi T(\phi ^{\prime })^{2}\sum_{\omega >0}\left( D_{s}\left[ \Im 
\frac{1}{\omega _{+}}\right] ^{2}\right.   \nonumber \\
&&\qquad \left. -\Re \frac{\alpha (\mathrm{sh}2q_{+}d_{f}+2q_{+}d_{f})}{%
4q_{+}\Omega_{+}^{2}}-\frac{D_{f}}{2h_f}\Im \frac{1}{\omega _{+}}%
\right) .  \nonumber
\end{eqnarray}%
where $\omega _{+}=|\omega |+\Gamma _{+}$.

In the limit of thin F layer ($q_+ d_f\rightarrow 0$) and large transparency
of S/F interface ($\gamma_B\rightarrow 0$), we note that $%
\Gamma_+\simeq\alpha q_+^2d_f\simeq (\eta_f/\eta_s)h_f$, and it is easily
checked that the above formulas for $1/2m$ and 
$U(x)$ reduce to Eq.~(\ref{coef-schrod}) from Sec.~\ref{sec2}, 
when $\eta_f\ll\eta_s$.

For large F films and transparent S/F interface, we find that the critical
temperature is given by Eq.~(\ref{eq:Tc-thick}), where $\Gamma _{+}=(1+i)h$
and $h=\alpha \sqrt{h_{f}/D_{f}}$: $\Gamma _{+}$ can be interpreted as a
combination of both exchange field and spin-flip terms with equal weight. In
such case, it is known that transition into uniform state is of the second
order.\cite{tollis} The coefficients of Eq.~(\ref{coef:thick}) also simplify
to the form 
\begin{eqnarray}
E &=&-\ln \frac{T}{T_{c}},\qquad \frac{1}{2m}=\pi TD_{s}\Re \sum_{\omega >0}%
\frac{1}{[\omega +(1+i)h]^{2}}  \nonumber \\
U(x) &=&-\pi TD_{s}(\phi ^{\prime })^{2}\sum_{\omega >0}\left[ \Im \frac{1}{%
\omega +(1+i)h}\right] ^{2}.\qquad 
\end{eqnarray}%
Let us note that effective exchange field and spin-flip parameter scale 
is $h\sim (\xi _{s}/d_{s})\sqrt{h_{f}T_{c0}}$ if S and F layers
have comparable diffusion constants and conductivities. When $d_{s}\ll \xi
_{s}$, as we assumed from beginning, this leads to $h\gg T_{c0}
$, and therefore to complete superconductivity suppression. However, we
expect that our results hold qualitatively in more general case $\xi
_{s}\lesssim d_{s}$ when superconductivity is not completely suppressed.

\end{document}